\documentclass[aps,prl,preprint,superscriptaddress]{revtex4-1}
\usepackage{graphicx}% Include figure files
\usepackage{dcolumn}% Align table columns on decimal point
\usepackage{bm}% bold math
\usepackage[latin1]{inputenc}
\begin{document}

%\preprint{}

\title{Measurement of the surface susceptibility and the surface conductivity of atomically thin MoS$\rm_2$ by spectroscopic ellipsometry}

\author{Gaurav Jayaswal}
\affiliation{King Abdullah University of Science and Technology, Computer, Electrical, and Mathematical Sciences and Engineering Division, Thuwal 23955-6900, Saudi Arabia}

\author{Zhenyu Dai}
\affiliation{King Abdullah University of Science and Technology, Physical Science and Engineering, Thuwal 23955-6900, Saudi Arabia}

\author{Xixiang Zhang}
\affiliation{King Abdullah University of Science and Technology, Physical Science and Engineering, Thuwal 23955-6900, Saudi Arabia}

\author{Mirko Bagnarol}
\affiliation{Dipartimento di Fisica e Astronomia G. Galilei, Universit$\grave{a}$ degli studi di Padova, via Marzolo 8, 35131 Padova, Italy}

\author{Alessandro Martucci}
\affiliation{Dipartimento di Ingegneria Industriale, Universit$\rm \grave{a}$ degli studi di Padova, via Marzolo 9, 35131 Padova, Italy}

\author{Michele Merano}
\email[]{michele.merano@unipd.it}
\affiliation{Dipartimento di Fisica e Astronomia G. Galilei, Universit$\grave{a}$ degli studi di Padova, via Marzolo 8, 35131 Padova, Italy}


\date{\today}

\begin{abstract}
We show how to correctly extract from the ellipsometric data the surface susceptibility and the surface conductivity that describe the optical properties of monolayer MoS$\rm_2$. Theoretically, these parameters stem from modelling a single-layer two-dimensional crystal as a surface current, a truly two-dimensional model. Currently experimental practice is to consider this model equivalent to a homogeneous slab with an effective thickness given by the interlayer spacing of the exfoliating bulk material. We prove that the error in the evaluation of the surface susceptibility of monolayer MoS$\rm_2$, owing to the use of the slab model, is at least 10\% or greater, a significant discrepancy in the determination of the optical properties of this material. 
\end{abstract}

\maketitle

In 2004, Novoselov and co-workers discovered that a variety of two-dimensional (2D) crystals can be mechanically exfoliated from a bulk precursor \cite{Novoselov2004,  Novoselov2005}. This promoted intense research in the physical properties of this new class of materials. These single-layer atomic crystals are stable under ambient conditions, exhibit high crystal quality, and they appear continuous on a macroscopic scale \cite{Novoselov2005}. Two-dimensional materials have diverse electronic properties, ranging from insulating hexagonal BN \cite{Blake2011} and semiconducting transition-metal dichalcogenides \cite{Heinz2010}, to semi-metallic graphene \cite{Novoselov2004}. In addition their optical properties are exceptional: their strong optical contrast is useful in microfabrication \cite{Blake2007, Kis11}, they support both transverse electric and transverse magnetic surface modes \cite{Eberlein08, Ziegler07, Merano16SW}, their second order nonlinear optical response depends on their crystal orientation \cite{Zhao13, Heinz13, Merano216}, the fine-structure constant determines the absorption of graphene \cite{Nair2008}, retardation-field effects and the radiation-reaction electric field play a relevant role in their optical response \cite{Luca16, Merano17}.  
  
This diversity in optical properties causes us to overlook very important subtleties in their physical description. Optical experiments have been interpreted by modelling all the 2D materials as homogeneous slabs, with an effective thickness of the order of the interlayer spacing of the original exfoliating solid \cite{Blake2007, Kravets2010}. There are two reasons for this. The first, atomic force microscopes can indeed measure the thickness of these materials \cite{Novoselov2005}. This can be confused as an experimental confirmation of the slab model. The second, exfoliated 2D crystals can be single-layer or multi-layer and the slab model can easily be extended to treat multi-layer crystals.

A recent paper \cite{Merano16} fitted optical-contrast \cite{Blake2007}, ellipsometry \cite{Kravets2010} and absorption \cite{Nair2008} experiments on graphene by modelling a 2D crystal as a zero-thickness interface with a complex surface conductivity \cite{Hanson08, Pershoguba07, Galina15} and compared the results with those for the slab model. A chi-squared test on the reported fits rejected the slab model at the 0.1\% significance level or lower, while it was consistent with the surface-conductivity model. 

In the case of graphene a constant value for the refractive index has been generally assumed \cite{Blake2007, Kravets2010, Merano16} and this is not necessarily implied by the choice of the slab model. The question then arises as to what happens if we allow the refractive index of the slab in the optical spectrum vary freely.   

In the family of two-dimensional crystals, monolayer transition metal dichalcogenides such as MoS$\rm_2$, MoSe$\rm_2$, WS$\rm_2$ and WSe$\rm_2$ have particular interesting optical properties \cite{Heinz2014}. They are direct band semiconductors \cite{Heinz2010, Heinz2014}, while their bulk precursors have got an indirect band gap. Due to this property their optical constants change significantly in the visible spectrum. The monolayer MoS$\rm_2$ was the first one to be experimentally addressed \cite{Heinz2010} and it is by far the most studied single-layer crystal in this family.

Among the several experimental investigations of the linear optical response of a single-layer MoS$\rm _2$. Li et al. \cite{Heinz2014} use the slab model to determine a complex bulk dielectric function $\epsilon =\epsilon_1 - i\epsilon_2$ and a related complex bulk optical conductivity $\sigma_b=i\epsilon_0 \omega(\epsilon_1-1-i\epsilon_2)$ where $\epsilon_0$ is the vacuum permittivity and $\omega$ is the angular frequency of the light. They then express $\chi$ and $\sigma$ by simply multiplying the corresponding bulk quantities by an effective thickness equal to the interlayer spacing $d$ of the exfoliating bulk material  
\begin{eqnarray}
\label{sheet conductivity}
\sigma= d\epsilon_0 \omega \epsilon_2; \qquad \chi=d(\epsilon_1-1)
\end{eqnarray}
The authors claim that this is equivalent to treating the single layer MoS$\rm_2$ as a 2D layer with a sheet conductivity.

\begin{figure}
\includegraphics{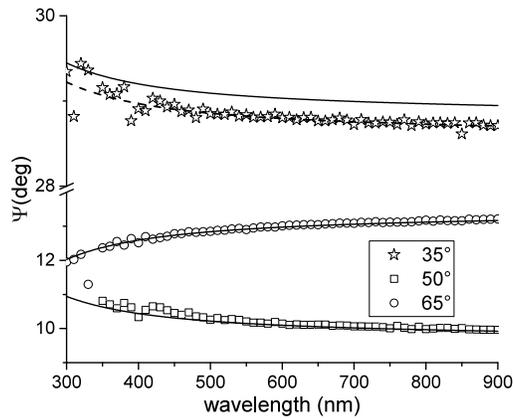}
\caption{\label{} Characterization of the substrate. Dots: experimental data. Solid lines: fits for a semi-infinite BK7 glass with n($\lambda$) provided by the Sellmeier equation. Dash line: fit at 35° for a semi-infinite BK7 glass with n($\lambda$)-0.020.}
\end{figure}

Morozov et al. \cite{Morozov15} publish both dielectric constants and optical conductivities. The dielectric constants are obtained starting from the slab model while the optical conductivities are obtained starting from the surface-current model. The two approaches are presented as equivalent; they claim: (A closer look at the boundary conditions for Maxwell equations suggests that the response of a very thin film to an electromagnetic field can equivalently be characterized by an electric field-induced surface current.) The results are presented without any further comparison in between the two approaches.

Shen et al. \cite{Shen13} report the real part of the surface conductivity in the THz range and the complex dielectric constant in the optical spectrum. They fit their ellipsometric spectra with a: (four-medium optical model, consisting of a semi-infinite substrate / bulk film / surface roughness / air ambient structure.) Their approach is equivalent to ref. \cite{Heinz2014}.

Many other authors determined the optical properties of a monolayer MoS$\rm_2$ \cite{Li14, Liu14, Funke16, Park14, Park216, Dai15, Duesberg14} They all use the slab model and describe the crystal as in \cite{Heinz2014}.

In this paper, we report ellipsometric data for a single-layer MoS$\rm_2$. The surface-current model is used to extract both $\chi$ and $\sigma$. Then these results are compared with those obtained applying the slab model plus equation (1). The different estimation due to the two models is finally discussed in terms of the experimental uncertainty of our measurements.

We determine $\chi$ and $\sigma$ by fitting the ellipsometric measurements of a single-layer MoS$\rm_2$ deposited on a transparent substrate. The precision of extracting the optical constants of a monolayer MoS$\rm_2$ can be substantially increased in this case because the ellipsometric $\Delta$ response of a dielectric substrate is trivially 180° for an angle of incidence smaller than the Brewster's angle ($\theta_B$) and 0° otherwise. Thus any ellipsometric phase variation should be produced by the MoS$\rm_2$ layer.

Spectroscopic ellipsometric measurements are performed using a VASE ellipsometer (J. A. Wollam) in ambient conditions at room temperature, for 3 angles of incidence (35°, 50°, 65°) in the spectral range 300 nm $\rm \le$ $\lambda$ $\rm \le$ 900 nm that encompasses the entire visible spectrum ($\lambda$ is the wavelength of the incident light).

We have prepared one large-area (up to millimeters), poly-crystalline, continuous, single-layer MoS$\rm_2$ with chemical vapor deposition (CVD). A SiO$\rm_2$/Si substrate was used for the growth process before the transfer on a BK7 slide 1 mm thick. This large-sized sample facilitates the ellipsometric measurement. We ensure the reproducibility of the experimental data by varying the position and the angle of incidence of the incoming beam when the spectra are collected.

Backside reflections are a problem when using parallel plate thin transparent substrates. They alter the measurement of the optical properties of the 2D material deposited on the top. We taped the backside of the substrate to avoid this problem.

\begin{figure}
\includegraphics{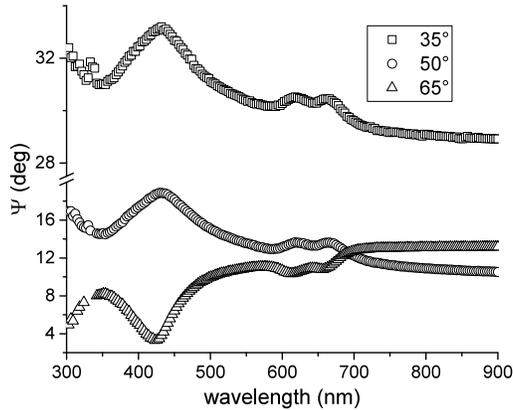}
\caption{\label{} Experimental data for the single-layer MoS$\rm_2$ on a BK7 substrate: ellipsometric parameter $\Psi$.}
\end{figure}

The first step of our analysis is the characterization of the substrate. At 50° and 65° angles of incidence the ellipsometric $\Psi$ parameter of the substrate is perfectly fitted by assuming a semi-infinite BK7 glass and the Sellmeier expression for its refractive index n($\lambda$). At 35° the best fit is obtained by considering the Sellmeier expression for n($\lambda$) minus a constant value of 0.020 (fig.1). The ellipsometric parameter $\Delta$ is dominated by noise. We measure the extinction coefficient of the substrate by an absorbance measurement (Jasco Spectrophotometer V-550) and we find that it is minor than $ 1.43\cdot10^{-5}$ across the spectral range that we analyze. We verified that such a low extinction coefficient does not influence the fit of the $\Psi$ parameter of the substrate nor the other experimental fits reported in this paper. 

Then we measure the sample. Figure 2 reports the experimental data for the ellipsometric parameter $\Psi$. The 2D crystal contributes a signal that is clearly appreciable in comparison with the substrate. It is interesting to note that the $\Psi$ of the sample has similarities with that of the substrate in its magnitude and in its dependence on the incident wavelength. Before $\theta_B$, $\Psi$ decreases with the wavelength (apart from the excitonic contribution of the sample). Above $\theta_B$, it increases with the wavelength. Also the ellipsometric parameter $\Delta$ (fig. 3) bears some similarity with a semi-infinite transparent substrate. Before $\theta_B$, it varies around 180°, after $\theta_B$ around 0°.

\begin{figure}
\includegraphics{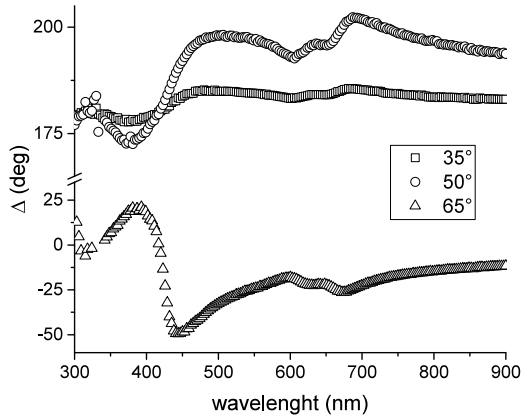}
\caption{\label{} Experimental data for the single-layer MoS$\rm_2$ on a BK7 substrate: ellipsometric parameter $\Delta$.}
\end{figure}

Figure 4 reports the most important result of the paper. The optical constants of a single-layer MoS$\rm _2$ are extracted from the experimental data using the surface-current model, a truly 2D model. Starting from $\Psi$ and $\Delta$, we obtain $\chi$ and $\sigma$ using the Fresnel coefficients of ref. \cite{Merano16} (formulas (6)). It is convenient to use a logarithmic scale for the $y$ axis of $\sigma$ because its value spans two order of magnitude. The good quality and the reproducibility of the experimental data is confirmed by the excellent superposition of the curves for $\chi$ and $\sigma$ at the three angles of incidence. Data for the three curves are collected every 3 nm at the same $\lambda$. For each $\lambda$ we can compute an average $\chi$ and $\sigma$. We can then compute the root mean square value and we obtain on average 0.35 nm for $\chi$ and $\rm 6.5\cdot10^{-6}$ $\Omega^{-1} $ for $\sigma$. The value of $\sigma$ is very sensitive to the refractive index of the substrate (at 900 nm a variation of n by 0.005 changes $\sigma$ of $\rm 7.5\cdot10^{-6}$ $\Omega^{-1}$). This is why it was important to well characterize it. Remarkably in our analysis we do not require any hypothesis on the roughness of the substrate (that can be neglected) or the existence of a Cauchy sublayer that are sometimes invoked in the ellipsometric measurements of 2D materials \cite{Funke16, Kravets2010}.  

\begin{figure}
\includegraphics{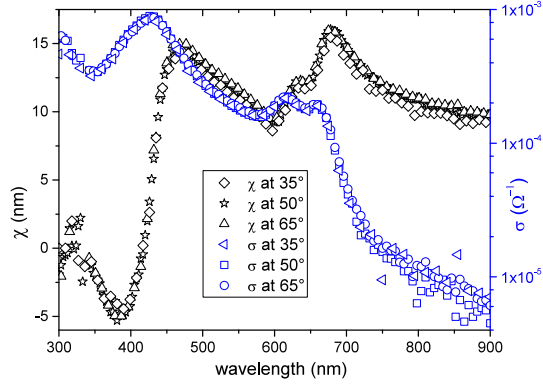}
\caption{\label{} Optical constants $\chi$ and $\sigma$ for a  single-layer MoS$\rm _2$ as a function of the incident wavelength. The superposition of the curves, extracted from data at three different angles of incidence, is excellent.}
\end{figure}

We now compare $\chi$ and $\sigma$ obtained with the surface-current model with those obtained with the slab model plus equation (1) and assuming a crystal thickness of 6.15 $\mathring{A}$ (i.e the interlayer spacing of the exfoliating bulk material) as it is usually done in literature. For clarity reasons, fig. 5 reports this analysis for the data collected at an angle of incidence of 65°. Anyway a similar analysis can be carried out for 35° and 50° with the same results. The average root mean square difference of the $\sigma$ extracted with the two models is $5.5\cdot10^{-6}$ $\Omega^{-1}$ i.e. it is practically within the experimental error of our data. The average root mean square difference of the $\chi$ extracted with the two models is 1.48 nm, clearly bigger than the experimental error reported above. This analysis is visually confirmed in fig. 5 where the respective curves are reported. 

\begin{figure}
\includegraphics{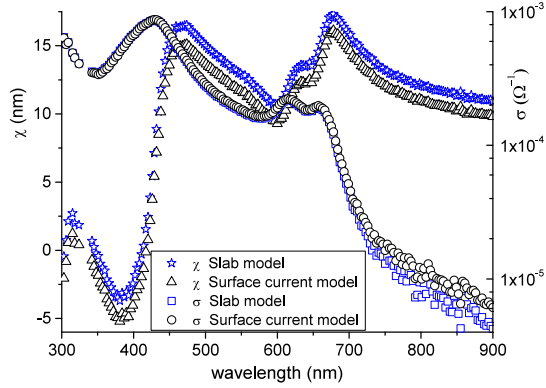}
\caption{\label{} Comparison of the optical constants of a single-layer $\rm MoS_2$ deduced from the surface-current model and the slab model. The two models give different results. Especially for $\chi$ this difference is greater than the experimental error, proving that the two models can not be considered equivalent.}
\end{figure}

As a final analysis we confirm once more that ellipsometry is a viable tool to ascertain if a 2D crystal is single-layer or not. Figure 6 compares the surface conductivity published in ref. \cite{Heinz2014} (fig. 3.a) with that measured by us. To allow comparison with \cite{Heinz2014}, we use a linear scale on the y-axis. The values of $\sigma$ for a single-layer $\rm MoS_2$ crystal in comparison with a two-layer crystal are much bigger than the different results given by the two models considered here. So even if ref. \cite{Heinz2014} uses the slab model, it is irrelevant for this specific analysis. With respect to the exfoliated sample of \cite{Heinz2014}, our CVD sample shows broader peaks, but it is clearly a single-layer crystal. 

In conclusion we have carried out spectroscopic ellipsometric measurements on a single-layer $\rm MoS_2$ in the spectral range 300 nm $\rm \le$ $\lambda$ $\rm \le$ 900 nm. The optical constants of the material, $\chi$ and $\sigma$, have been extracted via the surface-current model \cite{Merano16}, specific for a 2D physical system. It is interesting to compare these results with similar analysis done on graphene and single layer h-BN \cite{Merano16SW, Merano16}. These materials have optical parameters that are practically constant in the visible spectrum. The value of $\chi$ for these two materials is approximately 1nm i.e one order of magnitude smaller than the values spanned by the single-layer $\rm MoS_2$. The value of $\sigma$ for graphene is $\rm =6\cdot 10^{-5}$ $\Omega^{-1}$ while for h-BN it was only possible to put an upper limit of $\sigma \leq 2\cdot 10^{-6} \Omega^{-1}$. This is representative of the semi-metallic and the insulating character of these two materials. The $\sigma$ of the single-layer $\rm MoS_2$ is below $10^{-5} \Omega^{-1}$ in the infrared and bigger than $10^{-4} \Omega^{-1}$ starting from the first excitonic resonance.  

The results published on graphene \cite{Merano16} state the superiority of the surface-current model versus the slab model if a constant value for the refractive index is assumed \cite{Blake2007, Kravets2010}. Here we have considered a 2D material where the optical constants vary appreciably in the visible spectrum. We have shown that, contrarily to what is claimed in literature \cite{Heinz2014, Morozov15, Shen13, Li14, Liu14, Funke16, Park14, Park216, Dai15, Duesberg14}, the slab model and the surface-current model are not equivalent even if in the analysis of the experimental data we allow the refractive index of the hypothetical 3D slab vary freely. The error due to the slab model in the evaluation of the surface susceptibility of monolayer MoS$\rm_2$ is at least 10\% or greater, a significant discrepancy in the determination of the optical properties of this material.     

\begin{figure}
\includegraphics{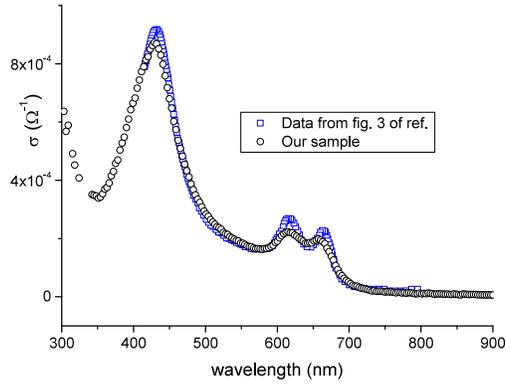} 
\caption{\label{} Our sample (circle dots) versus the exfoliated one (square dots) reported in ref \cite{Heinz2014}. The magnitude of $\sigma$ confirms that our $\rm MoS_2$  is single-layer.}
\end{figure}

\section*{Author Contributions}

M.M conceived the idea and wrote the paper. G. J. and Z. D prepared the sample under the supervision of X. Z. Measurements and data analysis were done by M. B., M. M. and A. M.

\bibliography{biblio}

\end{document}